\documentclass{PoS}
\usepackage{sidecap}

\title{Development of a Cherenkov Telescope for the Detection of Ultrahigh Energy Neutrinos with EUSO-SPB2 and POEMMA}

\ShortTitle{EUSO-SPB2 Cherenkov Telescope}
\author{\speaker{A.~Nepomuk Otte}$^a$, Eliza Gazda$^a$, Eleanor Judd$^b$, John F.\ Krizmanic$^c$,  Evgeny Kutzenzov$^d$, Oscar Romero Matamala$^a$, Patrick J. Reardon$^{de}$, and Lawrence Wiencke$^f$ -- JEM-EUSO and POEMMA Collaborations \footnote{for collaboration list see PoS(ICRC2019)1177} \\ 
\llap{$^a$}School of Physics \& Center for Relativistic Astrophysics, Georgia Institute of Technology,\\\ 837 State Street NW, Atlanta, GA 30332-0430, USA\\
\llap{$^b$} UC Berkley, Space Sciences Laboratory\\
\llap{$^c$} University of Maryland, Baltimore County\\
\llap{$^d$} The University of Alabama in Huntsville\\
\llap{$^e$} Center for Applied Optics\\
\llap{$^f$} Colorado School of Mines\\
E-mail: \email{otte@gatech.edu}}

\abstract{The detection of astrophysical neutrinos by IceCube and the potential to constrain source models of ultra-high energy cosmic rays provide the motivation to develop instruments for the observation of neutrinos above $10^7$\,GeV. Among the different techniques to detect ultra-high energy neutrinos is the Earth-skimming technique. It makes use of the fact that the tau produced in a tau neutrino interaction inside the Earth can emerge from the ground and initiate an upward-going particle shower in the atmosphere. The particle shower and thus the neutrino can be reconstructed by measuring the Cherenkov and radio emission from the shower particles. In this presentation, we discuss our ongoing development of a Cherenkov telescope for the detection of tau neutrinos, which is to be deployed on the Extreme Universe Space Observatory Super Pressure Balloon 2 (EUSO-SPB2) and is a precursor experiment for the proposed Probe of Extreme Multi-Messenger Astrophysics (POEMMA) mission. POEMMA aims at the detection of ultrahigh energy cosmic rays and ultrahigh energy neutrinos from low earth orbit. The 1\,m$^2$ Cherenkov telescope for EUSO-SPB2 will use silicon photomultipliers coupled to a 100 MS/s readout based on the ASIC for General Electronics for TPC`s (AGET) switch capacitor ring sampler. We present the optics, results from our studies to qualify the readout concept and the design of the mechanical integration of the photon detectors and the readout into the telescope.}

\FullConference{36th International Cosmic Ray Conference -ICRC2019-\\
		July 24th - August 1st, 2019\\
		Madison, WI, U.S.A.}

\begin{document}

\section{Introduction}

The development of the Cherenkov telescope presented here is motivated by the scientific potential of the ultrahigh energy (UHE) neutrino band, \emph{i.e.}\ neutrinos of astrophysical origin with energies above $10^7$\,GeV. If the measurement of the astrophysical neutrino flux detected by IceCube \cite{Aartsen2013} could be extended into the UHE band, it would shed light on the neutrino sources. Another science case for UHE neutrinos is the long-standing quest to  understand the composition and the sources of UHE cosmic rays (UHECRs), which is imprinted in the flux of UHE neutrinos that are produced when UHECRs interact with cosmic background radiation, see \emph{e.g.}\,\cite{2019arXiv190306714A,Anchordoqui2019}. 
The third case for UHE neutrinos is to test neutrino physics at the highest energies, which could potentially hint at new physics beyond the standard model, see \emph{e.g.}\,\cite{2019arXiv190700991B}. And last but hot least, UHE neutrinos are an important probe of beyond Standard Model physics, neutrino cross sections~\cite{Klein2013,2019arXiv190304333A}, and superheavy dark matter, see \emph{e.g.}~\cite{2012JCAP...10..043M}. 

UHE neutrinos are guaranteed to exist but thus far have not been observed. But several experiments employing different detection techniques have been proposed with the goal of a first UHE neutrino detection. One of these experiments is POEMMA, a satellite-borne instrument to measure orders-of-magnitude higher numbers of UHECRs than attained by ground-based observatories \cite{Olinto2017}. POEMMA aims to detect UHE neutrinos with the earth-skimming technique. In the earth-skimming technique, UHE neutrinos enter the Earth's surface at a small angle and interact after a few ten kilometers in rock, see \emph{e.g.}\,\cite{Fargion1999}. In case of a tau neutrino, a tau is produced, which continues to travel through the Earth. If it does not decay before, the tau emerges from the ground and decays in the atmosphere where a particle shower develops. The shower particles emit Cherenkov light, which is detected by POEMMA and used to reconstruct the tau neutrino.

Here we present the development of a Cherenkov telescope with a silicon-photomultiplier camera, which will be integrated onto the Extreme Universe Space Observatory Super Pressure Balloon 2 (EUSO-SPB2) as a precursor experiment to POEMMA \cite{Adams2017}. With EUSO-SPB2 we will investigate the feasibility to detect Cherenkov light from upward going tau showers and distinguish the Cherenkov signal from that of other background sources like fluctuations in the ambient photon field. 

\section{The EUSO-SPB2 payload}

\begin{SCfigure}[1.0][t]
\includegraphics[width=.6\textwidth]{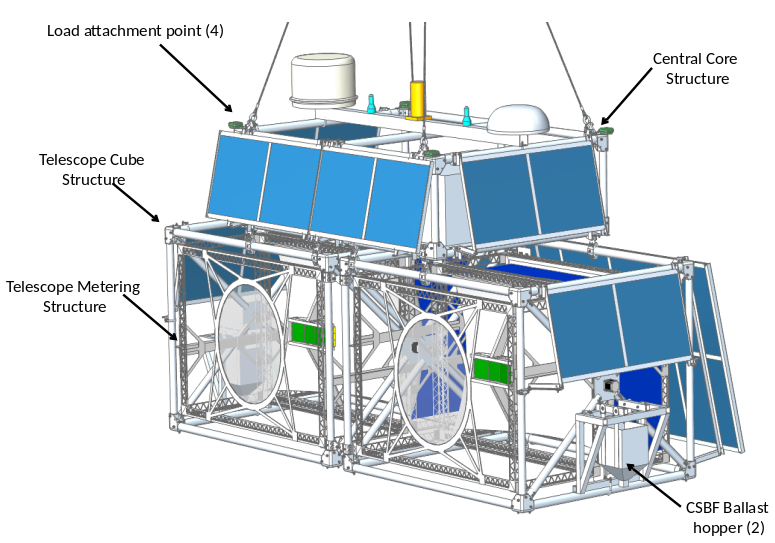}
\caption{Gondola with the fluorescence telescope and the Cherenkov telescope mounted on its bottom. The cameras of both telescopes are shown in green. The Cherenkov telescope will point at the Earth's limb during observation. The gondola hosts the batteries, computers, means of communication, and the remaining balloon infrastructure}
\label{fig:SPB2}
\end{SCfigure}

Figure \ref{fig:SPB2} shows the gondola and payload of the EUSO-SPB2 mission. The payload consists of two telescopes. One telescope is being developed to detect fluorescence light from UHECRs while the second one is the Cherenkov telescope, which is presented here. The two telescopes are powered with Li-phosphate batteries, which are charged through solar panels. The capacity of the batteries is sufficient to provide a 500\,W night time load and a 170\,W daytime load. The estimated weight of the two telescopes including cameras and data acquisition system is about 1 ton. The total weight of the gondola is 2.5 tons. An up-to-date status report of the EUSO-SPB2 mission and the fluorescence detector is presented in \cite{Wiencke2019}. The development of a SiPM based camera for the fluorescence detector is discussed in \cite{Painter2019}.

\section{Optical system}

\begin{figure}[b]
\centering
\includegraphics[width=.5\textwidth]{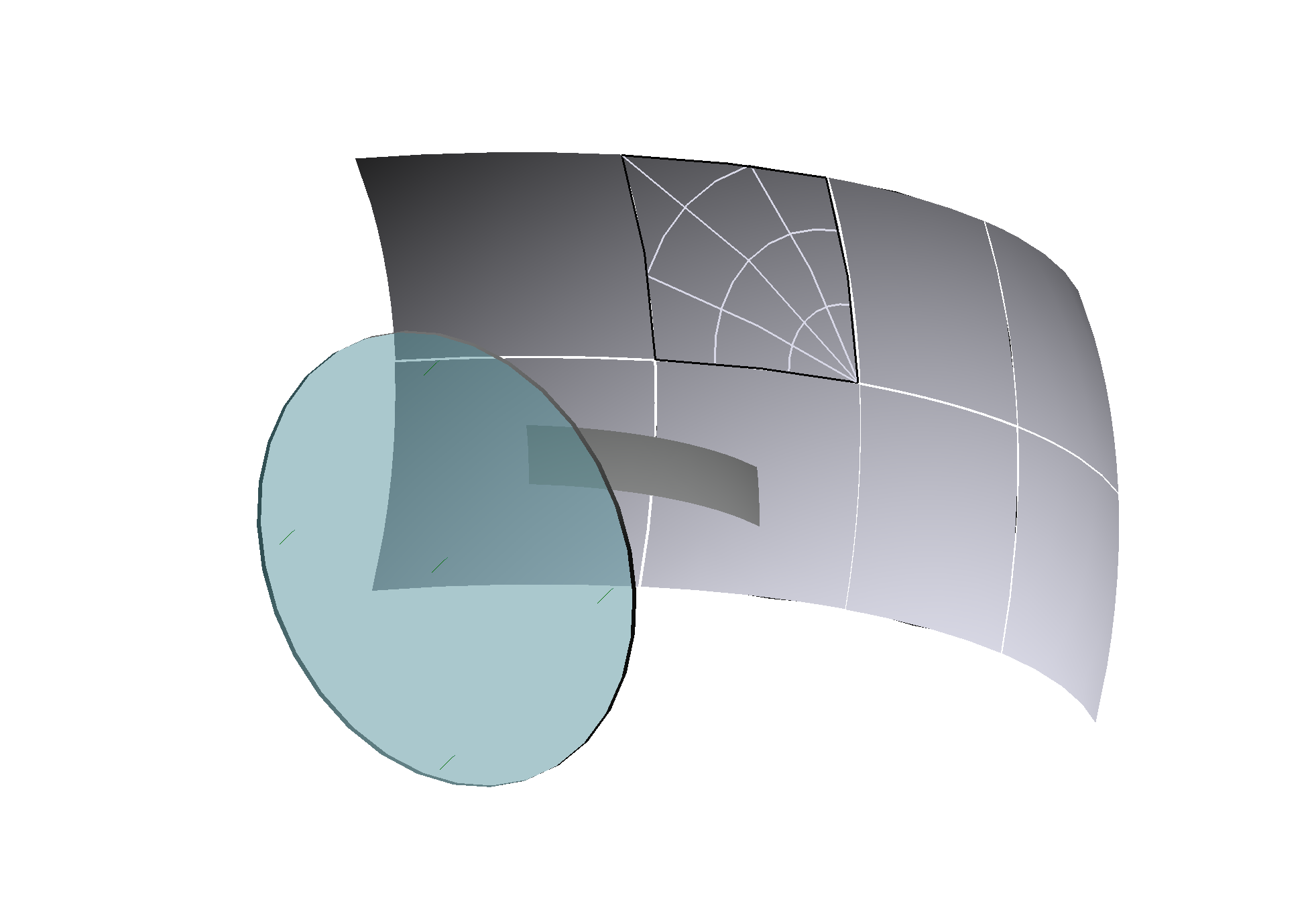}
\hfill
\includegraphics[width=.4\textwidth]{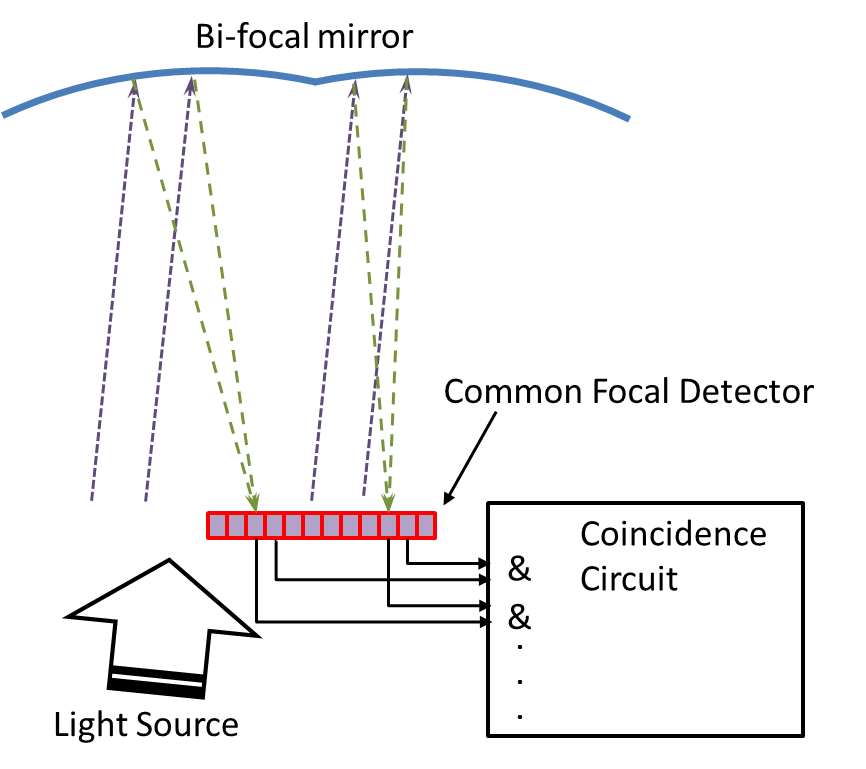}
\caption{Left: Schmidt catadioptric optics used for the EUSO-SPB2 telescopes. The optics consist of a PMMA corrector plate through which the light enters the system and is then reflected off the tessellated mirror surface consisting of eight mirror elements. The mirrors focus the light onto the curved focal plane, which is located between the corrector plate and the mirror. Right: The bifocal concept projects the image of a source onto two different locations in the camera. A coincidence trigger logic is formed between corresponding pixels in the camera.}
\label{fig:optics}
\end{figure}

The optics for both the Cherenkov and Fluorescence systems are based on a Schmidt catadioptric system. A CAD drawing of the optics is shown on the left side of Figure \ref{fig:optics}. The total size of the mirror surface is $2.1$\,m$\times1.1$\,m and consists of 8 identical segments. The system has an effective focal length of 860\,mm with a focal plane that is curved with a 850\,mm radius. Including alignment uncertainties and the optical quality of the mirrors, we expect that eighty percent of the image of a point source is contained within a circle of 5\,mm diameter. If the focal plane is populated with 3\,mm pixels, one pixel covers $0.2^\circ$ in the sky. The angular resolution is thus dominated by the size of the camera pixels. The optical design supports a focal plane covering a solid angle of approximately $5^\circ\times40^\circ$. Between $5^\circ\times5^\circ$ and $5^\circ\times10^\circ$ will be populated with photon detectors during the EUSO-SPB2 flight.

For the Cherenkov telescope we use a bi-focal concept -- see right panel in Figure \ref{fig:optics}. By tilting the optical axis of the upper and lower row of the telescope mirrors relative to each other, the image of an object appears twice in the focal plane with a well defined offset. We use a bi-focal optics because the air shower develops several hundred kilometers away and the shower axis points into the direction of the telescope. The Cherenkov signal of the air shower is, therefore, contained in only one pixel of the camera, which is difficult to discriminate from a random fluctuation in the ambient photon field. With the bi-focal optics, a coincidence can be formed between corresponding pixels substantially reducing the chance of triggering on a fluctuations and lowering the trigger threshold. 

\section{Camera}

\begin{SCfigure}
\centering
\includegraphics[width=.55\textwidth]{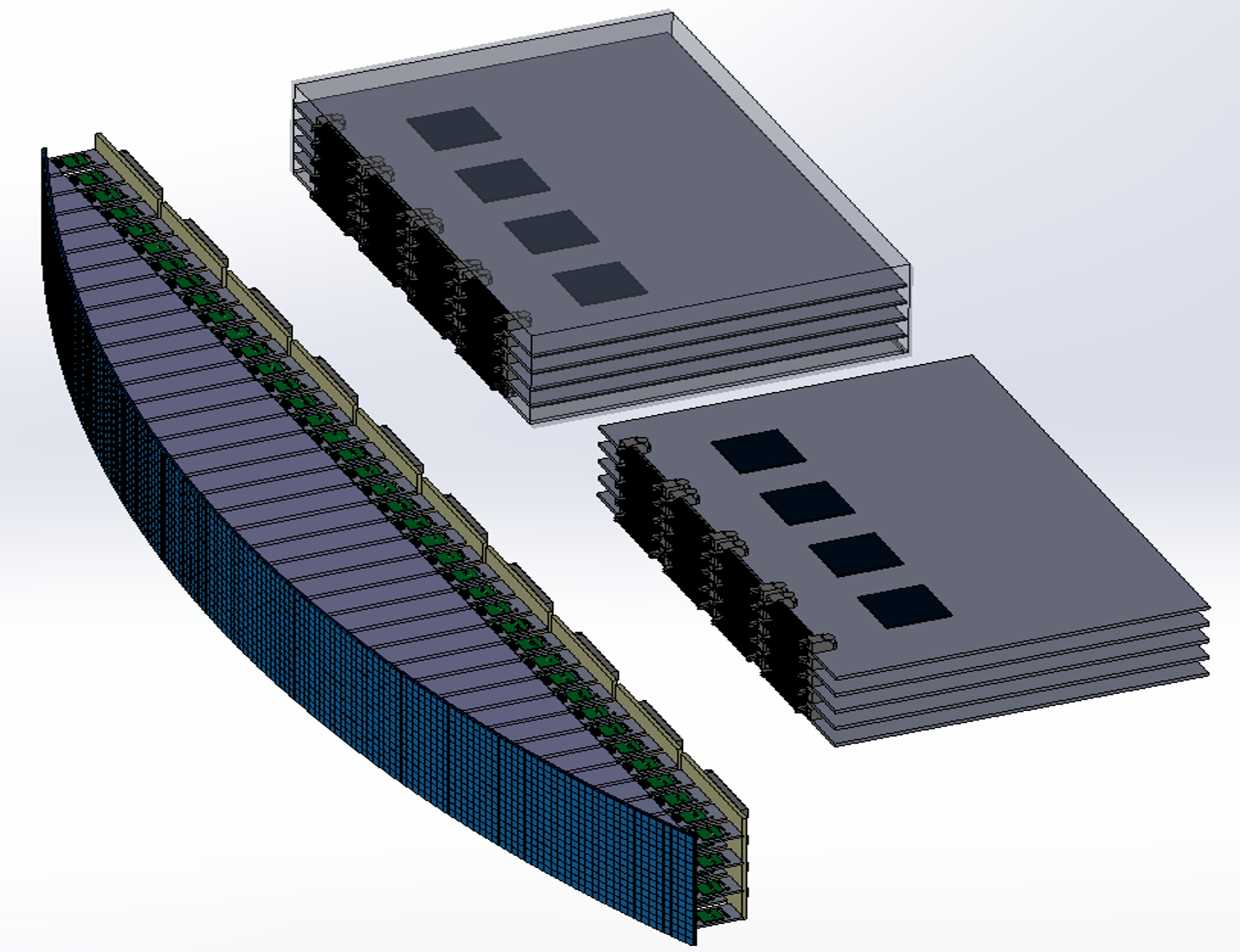}
\caption{CAD drawing of the camera developed for the Cherenkov telescope. In the configuration shown, the camera has 2560 pixel each with a size of $4.7$\,mm$\times4.7$\,mm. The silicon photomultipliers (blue) connect via interface boards of different sizes (grey) to the front-end electronics (green). The amplified signals are connected via cables (not shown) to AsAd boards (right) where they are being digitized.}
\label{fig:camera_structure}
\end{SCfigure}

Figure \ref{fig:camera_structure} shows a CAD drawing of a fully populated silicon photomultiplier camera with 2560 $4.7$\,mm$\times4.7$\,mm pixel , which is being developed for the Cherenkov telescope. Groups of 16 SiPMs are mounted onto one carrier board, which connects to an adapter printed circuit board (PCB). The dimensions of each adapter PCB are adjusted such that all SiPMs on one board are located as close as possible to their optimal position in the focal plane. The SiPM signals are routed via the adapter PCB to the Sensor Interface \& Amplifier Board (SIAB) where they are being amplified and shaped. Coming from the SIAB, the signals are connected via cables to the digitizer system. For EUSO-SPB2 only the central part of the camera is populated with 512 pixels. The power consumption of the partial populated camera is less than 100\,W.

\subsection{Photon detectors}
\begin{SCfigure}
    \centering
    \includegraphics[width=0.7\textwidth]{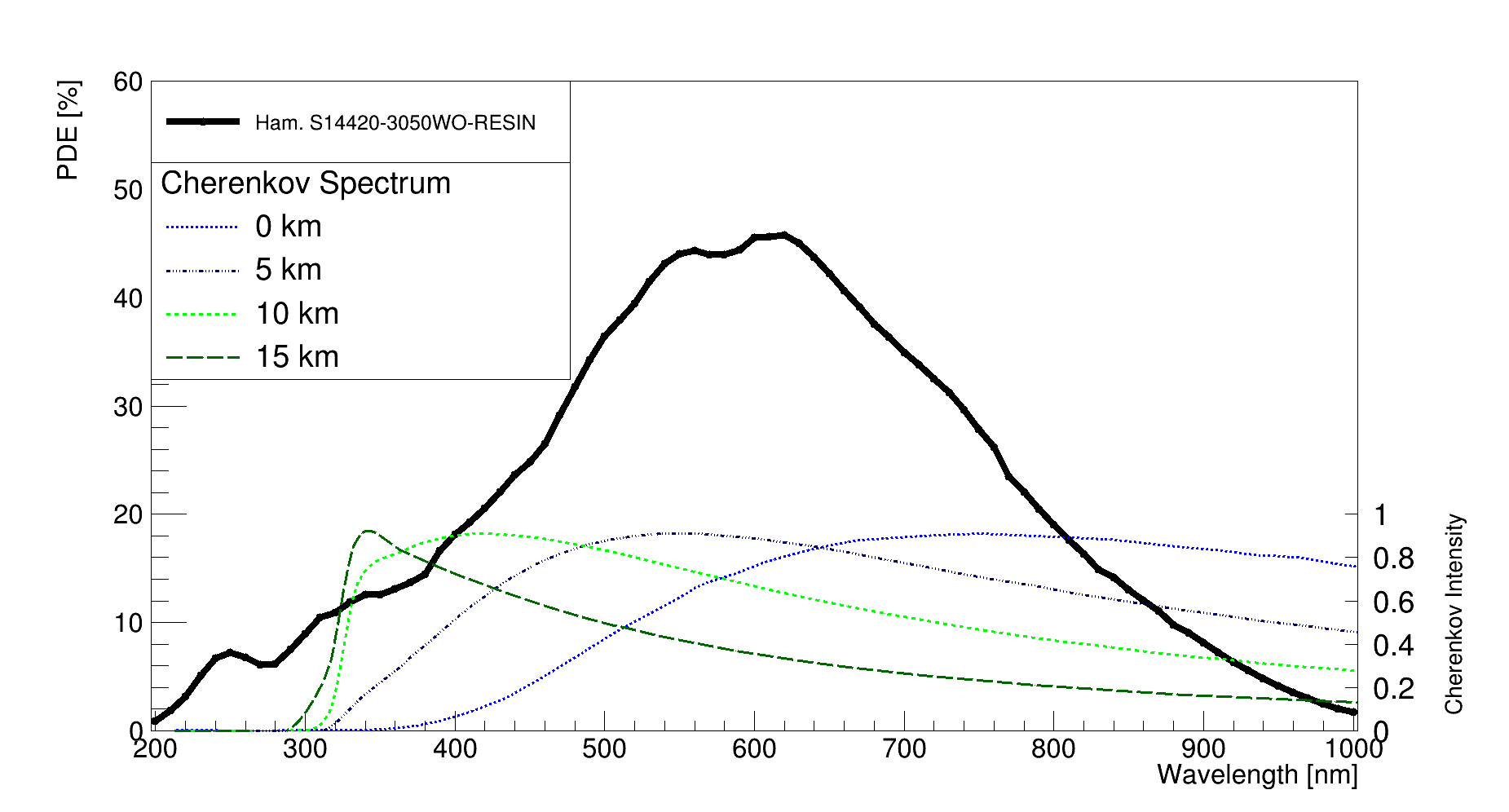}
    \caption{Photon detection efficiency of the red sensitive Hamamatsu SiPM S14420-3050WO-Resin together with normalized spectra of Cherenkov emission from air showers starting at different altitudes above ground. The apparatus used to measure the PDE is described in \cite{Otte2017}.}
    \label{fig:SiPMPDE}
\end{SCfigure}

The Cherenkov-telescope camera will use silicon photomultipliers (SiPMs). 
We have evaluated several SiPMs from FBK, Hamamatsu, and SensL; both blue sensitive and red sensitive ones. The spectral response of SiPMs is a good match to the Cherenkov spectrum, which peaks in the red after absorption in the atmosphere. As an example of the good match, we show in Figure \ref{fig:SiPMPDE} the photon detection efficiency (PDE) of the red sensitive Hamamatsu SiPM S14420-4050WO-Resin overlaid on top of simulated Cherenkov spectra expected at the entrance of the telescope. The Cherenkov spectra are normalized to the same peak amplitude. A decision which SiPM will be used in the camera will be made in the coming months once our SiPM evaluation is complete.

\subsection{Readout Electronics}
\begin{SCfigure}
    \centering
    \includegraphics[width=0.7\textwidth]{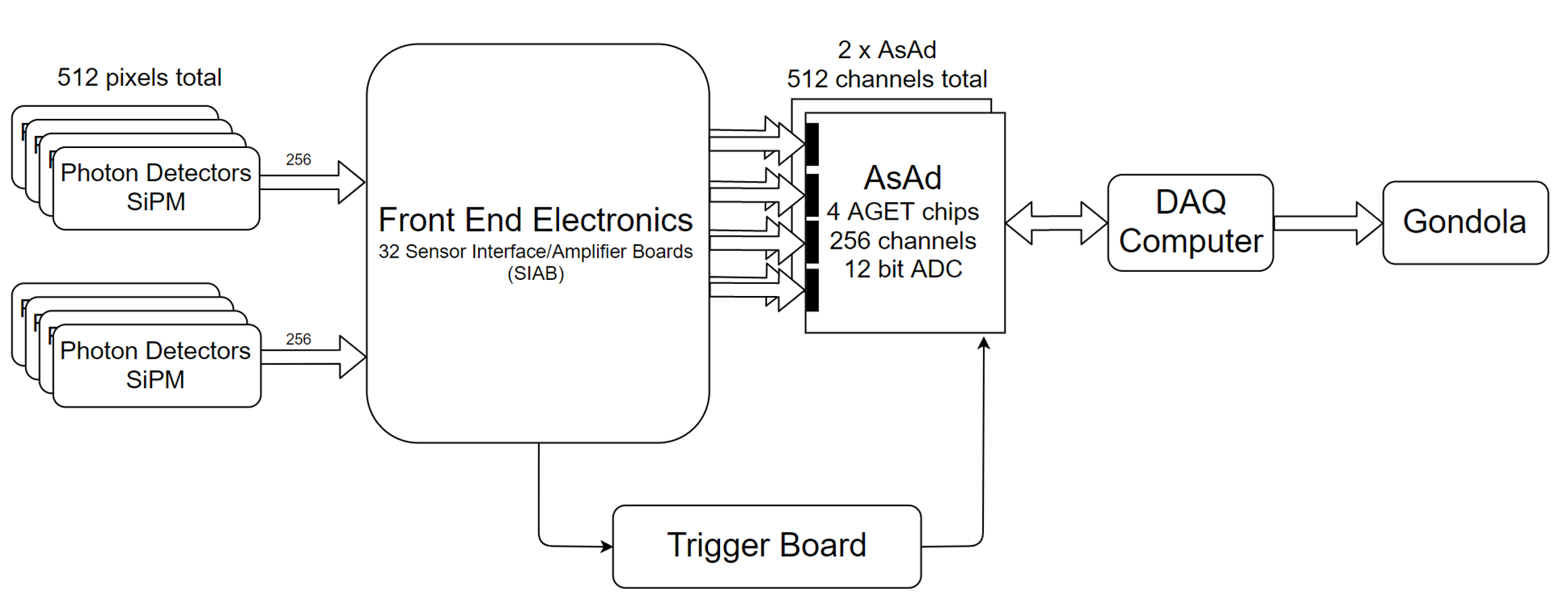}
    \caption{Signal chain used in the Cherenkov camera. The SiPM signals are amplified and shaped on the SIAB before digitized on the AsAd boards.}
    \label{fig:signal_chain}
\end{SCfigure}

The signal chain of the camera shows Figure \ref{fig:signal_chain}. It breaks down into the SIAB, the AGET digitizer system consisting of the AsAd board and a micro-TCA crate (not shown) with the concentration board (CoBo) and a network connection. A separate trigger board receives the discriminator signals from the SiPMs, forms a coincidence and sends a trigger signal to the CoBo.

\begin{SCfigure}
\centering
\includegraphics[width=.7
\textwidth]{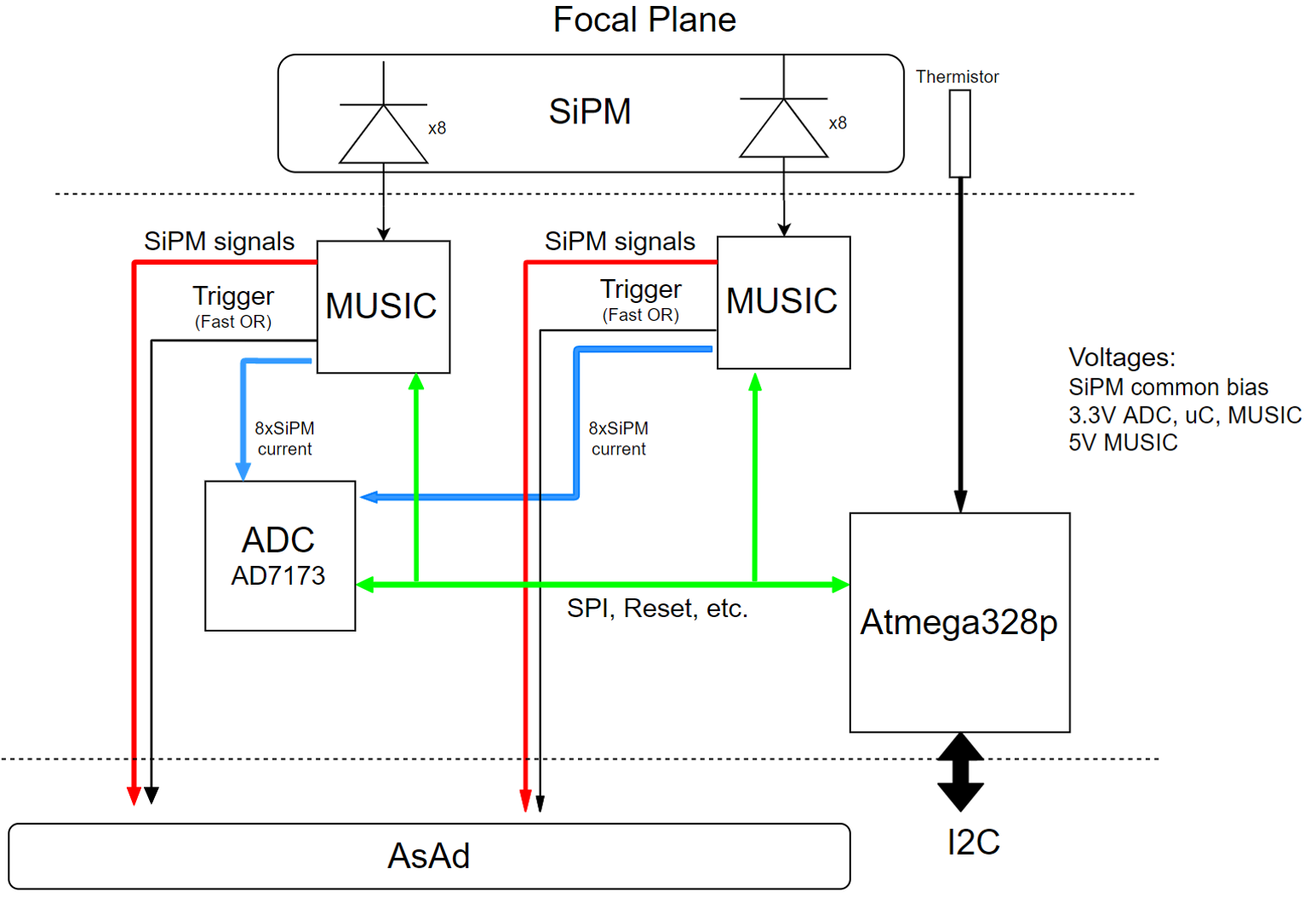}
\caption{Block diagram of the SIAB. See text for discussion.  }
\label{fig:SIAB_block}
\end{SCfigure}

\paragraph{SIAB} A block diagram of the SIAB is shown in Figure \ref{fig:SIAB_block}. Integrated on one SIAB are two Multiple Use SiPM Integrated Circuit (MUSIC) ASICs \cite{Gomez2016}, which amplify and shape the signals from 16 SiPMs. Additional housekeeping functionalities carried out by the MUSIC chip are adjusting the bias voltage of each SiPMs, turning channels on and off, and monitoring of the SiPM currents. The two MUSIC chips are controlled by an Atmega 328 microcontroller via an SPI interface. The SiPM currents are digitized with a 24-bit, low power AD7173 ADC, which connects to the microcontroller via SPI. The microcontroller also measures the temperature of the SiPM carrier board with a thermistor that is soldered on its back side. External communication with the microcontroller takes place via an I2C bus, which connects to all SIABs in the camera. The amplified and shaped SiPM signals connect to the AGET digitizer chips on the AsAd board via a backplane and a cable connection. The layout of the board is complete and a prototype is in production.  

\begin{SCfigure}[1.0][b]
    \centering
    \includegraphics[width=0.55\textwidth]{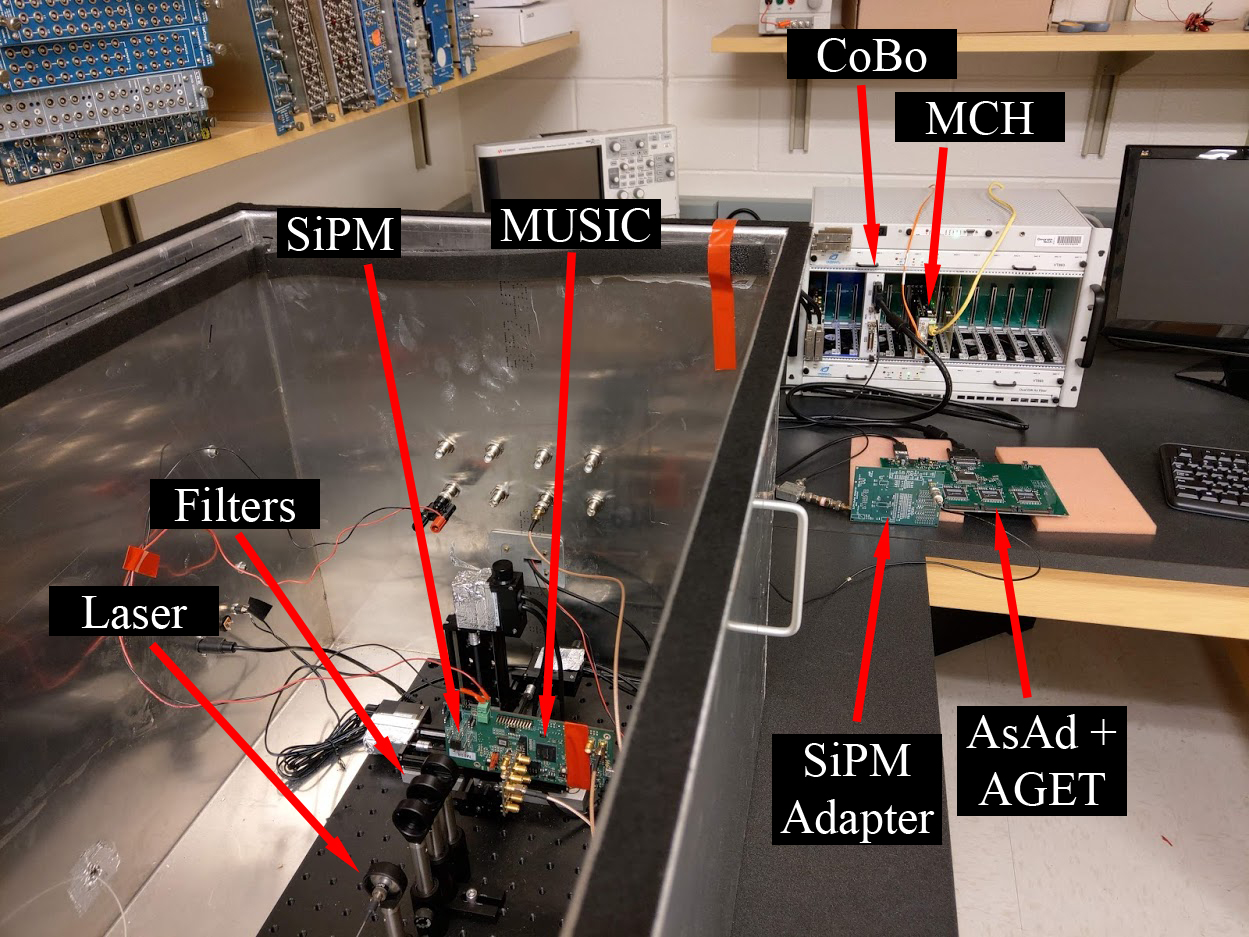}
    \caption{Test setup to evaluate the signal chain for the Cherenkov camera. Inside the dark box (on the left) the laser light propagates through calibrated neutral density filters before it illuminates an SiPM, which is inserted into the MUSIC evaluation board. Outside the box, the amplified SiPM signals connect to either an oscilloscope or the AsAd board with the AGET switch capacitor array chips. Also shown in the picture is the MicroTCA crate with the CoBo and MCH, which are discussed in the text.} 
    \label{fig:testsetup}
\end{SCfigure}

\begin{SCfigure}
    \centering
    \vspace*{-5ex}
    \includegraphics[width=0.55\textwidth]{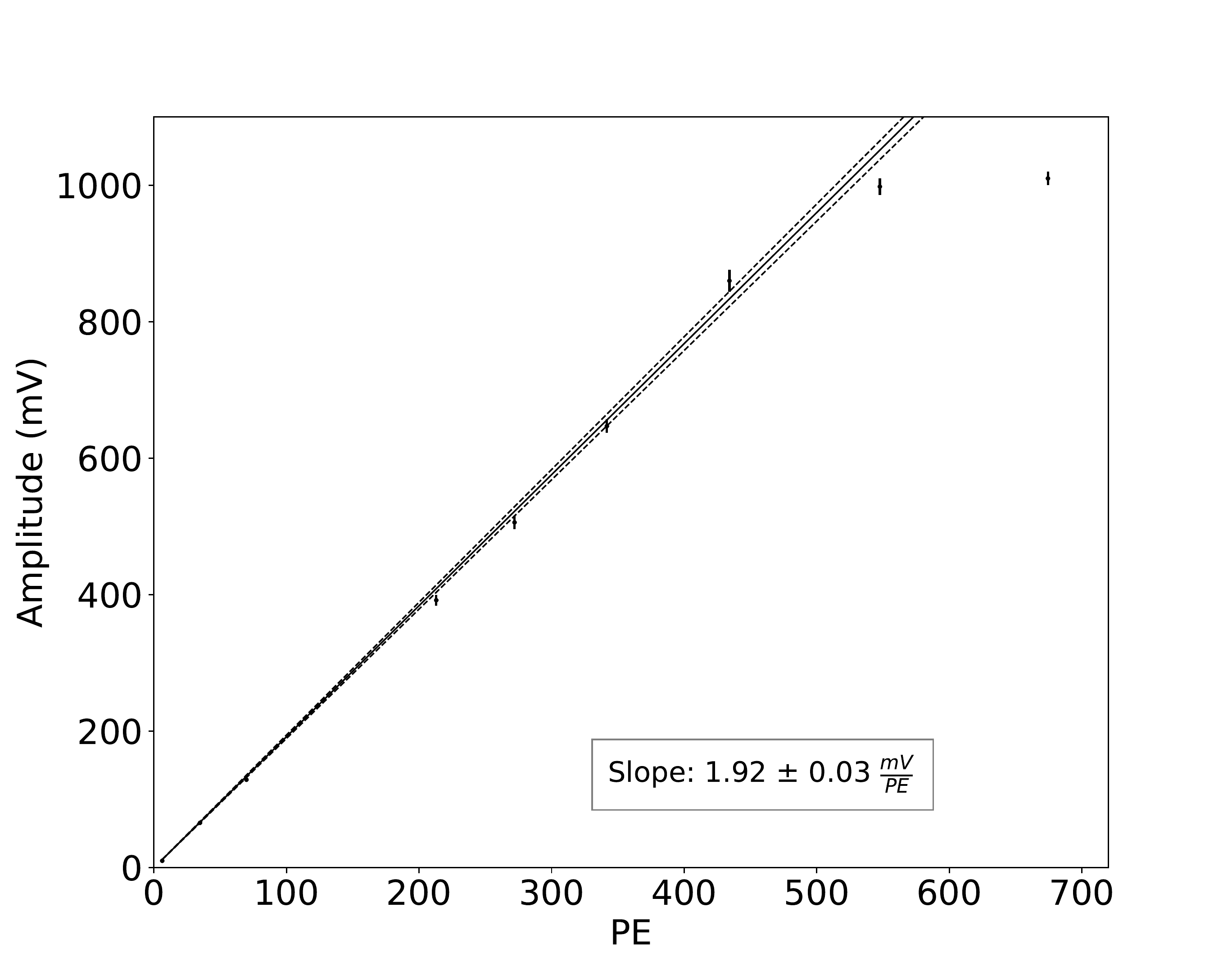}
    \caption{Amplitude response of the MUSIC chip to picosecond laser flashes recorded with a S14520-6050CN SiPM from Hamamatsu. The horizontal axis is calibrated in units of average photoelectrons (PE) detected by the SiPM per laser flash. The saturation above 500\,pe is due to the limitation of the MUSIC chip. The measurement was done at room temperature.}
\label{fig:MUSIC_linearity}
\end{SCfigure}

We have measured the performance of the MUSIC chip by flashing a Hamamatsu S14520-6050CN SiPM with a picosecond laser and varied the intensity of the flashes with calibrated neutral density filters. A picture of the setup shows Figure \ref{fig:testsetup}. The same setup is also used in the evaluation of the entire signal chain, which includes the digitizer. Figure \ref{fig:MUSIC_linearity} shows the amplitude response of the MUSIC chip measured with a 500\,MHz bandwidth oscilloscope. For this measurement and measurements, which include the AGET digitizer system, we have used a MUSIC evaluation board provided by the MUSIC chip developers. The bias voltage of the SiPM chosen for this measurement also maximizes the PDE of the SiPM. The response of the MUSIC chip is linear up to a signal equivalent to 500 detected photons (photoelectrons).

\paragraph{AGET digitizer} For signal digitization we use the ASIC for General Electronics for TPC's(AGET) \cite{Pollacco2018}. AGET is a 100 megasamples per second switch capacitor array (SCA) with a five microsecond buffer depth and 64 channels. We have configured the analog front-end of the AGET to bypass the internal shaping stages and couple the signals directly to the SCA preceding, inverting amplifier, which has gain 2 and a 750\,mV input dynamic range.

Four AGET chips are integrated on one ASIC Support \& Analog-Digital conversion (AsAd) board, which thus provides 256 channels. The system is commercially available. When the AsAd board receives a read-out command, it instructs the AGET chips to stop sampling and the stored analog samples are digitized with 12-bit resolution by an onboard ADC \cite{Pollacco2018}. The AsAd board is connected to the concentration board (CoBo), which resides inside a MicroTCA crate.  CoBo runs the real time operating system VxWorks and a firmware, which combines the data streams from the AsAd boards. CoBo creates a binary file with the data, which it sends to the DAQ computer via a MicroTCA Carrier Hub and a $10\,$Gb/s optical link. A separate $1$\,Gb/s is used to configure the AsAd board and CoBo. The DAQ computer is described in \cite{Scotti2019}.

\begin{SCfigure}
    \centering
    \vspace*{-4ex}
    \includegraphics[width=0.55\textwidth]{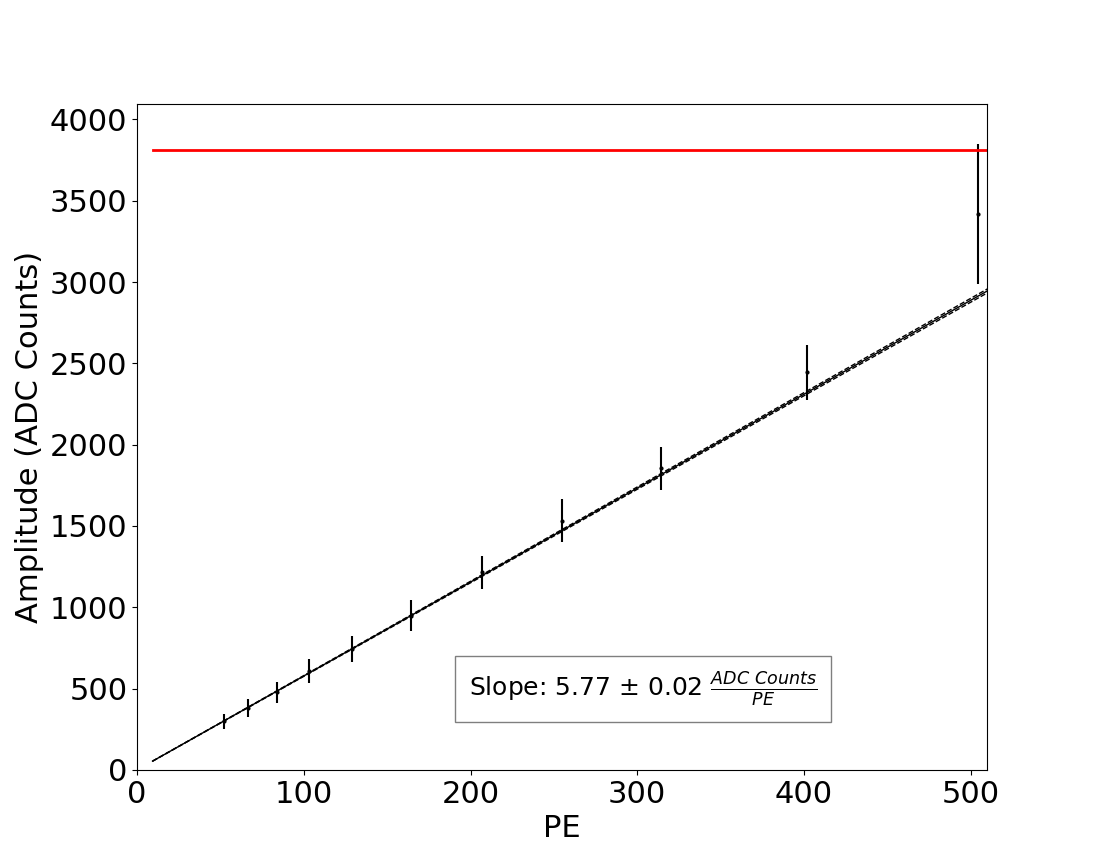}
    \caption{Amplitude response of the signal chain -- Hamamatsu S14520-6050CN SiPM, MUSIC chip, and AGET -- at room temperature to flashes of varying intensity from a picosecond laser. The error bars show the spread of the reconstructed amplitudes, while the data points show mean reconstructed amplitudes. The horizontal red line indicates the maximum value that can be digitized. The black line shows the fit of a linear function to the first six data points. The horizontal axis is calibrated in photoelectrons.}
    \label{fig:responsesignalchain}
\end{SCfigure}

Figure \ref{fig:responsesignalchain} shows the response of the AGET system to SiPM signals from the MUSIC chip. For this measurement, the setup in Figure \ref{fig:testsetup} was used and a 20\,MHz low-pass filter was inserted after the MUSIC evaluation board. The filter slows the picosecond laser signal to a rise time of 30\,ns emulating a realistic Cherenkov signal, that is spread out over several nanoseconds. The response of the system is linear to signals with amplitudes equivalent to about 400 photoelectrons, which is equivalent to the dynamic range of the MUSIC chip.

\paragraph{Trigger} The trigger logic is implemented in a Spartan-6 gate array, which receives the discriminator signals from all the MUSIC chips. The logic forms a coincidence between corresponding pixels in the focal plane that are expected to have correlated signals due to the bifocal optics. When a coincidence is found within a gate of 100\,ns, a trigger signal is sent to the CoBo board, which initiates the readout of the AGET chips.

\section{Discussion and Outlook}

We are developing a Cherenkov telescope for the EUSO-SPB2 mission. The MUSIC chip and the AGET system meet the requirements for EUSO-SPB2. Several SiPMs have been tested with spectral responses that are a good match to the expected Cherenkov spectrum. The tests of the SiPMs and the different camera components are being finalized this year, which includes thermal and vacuum testing. The camera construction is planned to begin in the first half of 2020. This research is supported by NASA grant 80NSSC19K0627.

\bibliographystyle{pos}
\bibliography{references}

\begin{thebibliography}{10}
\providecommand{\url}[1]{\texttt{#1}}
\providecommand{\urlprefix}{URL }
\providecommand{\eprint}[2][]{\url{#2}}

\bibitem{Aartsen2013}
M.~G. Aartsen, et~al., \emph{{Evidence for High-Energy Extraterrestrial
  Neutrinos at the IceCube Detector}}, \emph{Science} \textbf{342}~(6161)
  (2013) 1242856.

\bibitem{2019arXiv190306714A}
R.~Alves~Batista, et~al., \emph{{Open Questions in Cosmic-Ray Research at
  Ultrahigh Energies}} arXiv:1903.06714.

\bibitem{Anchordoqui2019}
L.~A. Anchordoqui, \emph{{Ultra-high-energy cosmic rays}}, \emph{Physics
  Reports} \textbf{801} (2019) 1.

\bibitem{2019arXiv190700991B}
P.~S. Bhupal~Dev, et~al., \emph{{Neutrino Non-Standard Interactions: A Status
  Report}} arXiv:1907.00991.

\bibitem{Klein2013}
S.~R. Klein et~al., \emph{{Neutrino Absorption in the Earth, Neutrino
  Cross-Sections, and New Physics}} arXiv:1304.4891.

\bibitem{2019arXiv190304333A}
M.~Ackermann, et~al., \emph{{Fundamental Physics with High-Energy Cosmic
  Neutrinos}} arXiv:1903.04333.

\bibitem{2012JCAP...10..043M}
K.~Murase et~al., \emph{{Constraining very heavy dark matter using diffuse
  backgrounds of neutrinos and cascaded gamma rays}}, \emph{Journal of
  Cosmology and Astroparticle Physics} \textbf{2012}~(10) (2012) 43.

\bibitem{Olinto2017}
A.~V. Olinto, et~al., \emph{{POEMMA: Probe Of Extreme Multi-Messenger
  Astrophysics}}, in \emph{35th International Cosmic Ray Conference} Busan,
  Korea2017 arXiv:1708.07599.

\bibitem{Fargion1999}
D.~Fargion, et~al., \emph{{Horizontal Tau air showers from mountains in deep
  valley. Traces of UHECR neutrino tau}} arXiv:astro--ph/9906450.

\bibitem{Adams2017}
J.~H. Adams, et~al., \emph{{White paper on EUSO-SPB2}} arXiv:1703.04513.

\bibitem{Wiencke2019}
L.~Wiencke, \emph{{The Extreme Universe Space Observatory on a Super-Pressure
  Balloon II Mission}}, in \emph{36th International Cosmic Ray Conference},
  2019 PoS (ICRC2019) 466.

\bibitem{Painter2019}
W.~Painter, et~al., \emph{{Silicon Photomultipliers for Orbital Ultra High
  Energy Cosmic Ray Observation}}, in \emph{36th International Cosmic Ray
  Conference}, 2019 PoS (ICRC2019) 285.

\bibitem{Otte2017}
A.~N. Otte, et~al., \emph{{Characterization of Three High Efficiency and Blue
  Sensitive Silicon Photomultipliers}}, \emph{NIM A} \textbf{846} (2016) 106.

\bibitem{Gomez2016}
S.~G{\'{o}}mez, et~al., \emph{{MUSIC: An 8 channel readout ASIC for SiPM
  arrays}} International Society for Optics and Photonics2016 98990G.

\bibitem{Pollacco2018}
E.~Pollacco, et~al., \emph{{GET: A generic electronics system for TPCs and
  nuclear physics instrumentation}}, \emph{NIM A} \textbf{887} (2018) 81.

\bibitem{Scotti2019}
V.~Scotti, \emph{{36th International Cosmic Ray Conference}}, in \emph{36th
  International Cosmic Ray Conference}, 2019 PoS (ICRC2019) 420.

\end{thebibliography}
%

\end{document}